\documentclass[preprint,12pt,authoryear]{elsarticle}

\journal{Journal of Neuroscience Methods}

\usepackage{wrapfig}    
\usepackage{epsfig,psfrag}
\usepackage{latexsym}
\usepackage{amssymb,amsmath}
\usepackage{tikz,pgf}
\usepackage{subfig}
\usepackage{float}
\usepackage{stfloats}
\usepackage{graphicx}
\usepackage{epstopdf}
\usepackage{color}

\usepackage{url}
\usepackage{hyperref}
\hypersetup{colorlinks,%
citecolor=black,%
filecolor=black,%
linkcolor=black,%
urlcolor=black}%

\usepackage[normalem]{ulem}

\usepackage{setspace}

\usepackage[singlelinecheck=false]{caption}

\date{}

\begin{document}
\begin{frontmatter}
\title{Statistical Model for Dynamically-Changing Correlation Matrices with Application to  Brain Connectivity}

\author[1]{Shih-Gu Huang}
\author[2]{S. Balqis Samdin}
\author[3,2]{Chee-Ming Ting}
\author[2]{Hernando Ombao}
\author[1]{Moo K. Chung\corref{mycorrespondingauthor}}
\cortext[mycorrespondingauthor]{Corresponding author}
\ead{mkchung@wisc.edu}
\address[1]{Department of Biostatistics and Medical Informatics, University of Wisconsin, Madison, WI 53706, USA}
\address[2]{Statistics Program, King Abdullah University of Science and Technology (KAUST), Thuwal 23955, Saudi Arabia}
\address[3]{School of Biomedical Engineering \& Health Sciences, Universiti Teknologi Malaysia, 81310 Skudai, Johor, Malaysia}

\begin{abstract}
{\em Background:} 
Recent studies have indicated that functional connectivity is dynamic even during rest.
A common approach to modeling the dynamic functional connectivity in whole-brain resting-state fMRI is to compute the correlation between anatomical regions  via sliding time windows.
However, the direct use of the sample correlation matrices is not reliable due to the image acquisition and processing noises in resting-sate fMRI.

{\em New method:} To overcome these limitations, we propose a new statistical model that smooths out the noise by exploiting  the geometric structure of correlation matrices. The dynamic correlation matrix is modeled as a linear combination of symmetric positive-definite matrices combined with cosine series representation. The resulting smoothed dynamic correlation matrices are clustered into disjoint brain connectivity states using the $k$-means clustering algorithm.

{\em Results:} The proposed model preserves the geometric structure of underlying physiological dynamic correlation, eliminates unwanted noise in  connectivity and obtains more accurate state spaces. The difference in the estimated dynamic connectivity states between males and females is identified.

{\em Comparison with existing methods:} We demonstrate that the proposed statistical model 
has less rapid state changes caused by noise and improves the  accuracy in identifying and discriminating different states.

{\em Conclusions:} We propose a new regression model  on dynamically changing correlation matrices that   provides better performance over existing windowed correlation and is more reliable for the modeling of dynamic connectivity.
\end{abstract}


\end{frontmatter}

\section{Introduction}

Findings of resting-state fMRI have revealed synchrony between spontaneous BOLD signal fluctuations in sets of distributed brain regions despite the absence of any explicit tasks.
Traditionally, brain functional connectivity between signals from  distinct brain regions  is often measured by the static correlation over the entire  scan duration. However, this simplification 
 of averaging over time cannot reveal the complex dynamics of the resting-state 
functional connectivity. 
Recent studies have suggested the dynamic changes in 
functional connectivity over time,
called \textit{dynamic functional connectivity}, 
even during rest \citep{Chang2010,Hutchison2013,Hutchison2015,Preti2017}.
The dynamic functional connectivity is also referred to as time-varying (functional) connectivity in literature \citep{calhoun2014chronnectome,lurie2018nature,thompson2018simulations}.

A common approach to modeling  the dynamic functional connectivity is through the sliding-window
method, where dynamic correlation  matrix is computed by the Pearson correlation over consecutive windowed segments of fMRI time series   over predefined brain  parcellation  \citep{keilholz2013dynamic,Hutchison2013,kucyi2014dynamic,Allen2014,Hutchison2015,Zalesky2015,Shakil2016,Hindriks2016}.
Crucial to subsequent inference is the estimation of the underlying
dynamic correlation matrix. Due to image acquisition and processing noises as well as the low signal-to-noise ratio in fMRI data, data smoothing is necessary.

In this paper, we  develop an approach that uses a canonical  series representation and hence will be robust to 
model misspecification and  have the ability to 
more accurately capture transient dynamics in connectivity. The proposed canonical  series representation is
related to the regression problem  on Riemannian manifolds.
The computations on Riemannian manifolds have been applied to various medical imaging applications such as  interpolation, regularization and estimation of diffusion tensors images \citep{arsigny.2006,fillard2007clinical,barmpoutis2007tensor,cheng2012efficient},
 shape modeling of corpus callosum
\citep{fletcher2013geodesic,hinkle2014intrinsic}, nonlinear mixed effects models on Cauchy deformation tensor  for analyzing longitudinal deformations in brain imaging \citep{kim2017riemannian}, and regression and classification of brain networks \citep{qiu.2015,wong2018riemannian}.

One can summarize the whole-brain dynamic functional connectivity into a smaller set of \textit{dynamic connectivity states}, defined as distinct transient connectivity patterns that repetitively occur throughout the resting-state scan \citep{Hutchison2015}.
The dynamic connectivity states are reliably observed across different subjects, groups, sessions and trials \citep{Yang2014,choe2017comparing,ombao2018statistical}.
In this paper, we apply the $k$-means clustering on the proposed smoothed correlation matrices to identify difference in dynamic connectivity states in resting-state fMRI between males and females. The $k$-means clustering on resting-state fMRI was introduced by \citep{Allen2014} and subsequently adopted by many others \citep{Damaraju2014,
Barttfeld2015,Hutchison2015,
Rashid2016,samdin2017unified,ting2018estimating}
to identify these recurring  dynamic connectivity states. The cluster centroids correspond to the connectivity patterns. It has been shown that additional summary metrics of the fluctuations in these clustering-derived states, such as the amount of time spent in specific states and the  transition probability between states, exhibit meaningful between-group variations such as age \citep{Hutchison2015,Marusak2017} and clinical status \citep{Damaraju2014,Rashid2016,Su2016,Barber2018}.

In this paper, we develop a robust  estimate of the dynamic correlation matrices, which serves as an input to the more refined state-space analysis.  The correlation matrices are modeled by a fixed set of  matrices whose  
matrix logarithms form an orthonormal basis in the  space of symmetric matrices.
The proposed statistical
model can preserve information of the underlying dynamic correlation and eliminate the rapidly changing noise in the connectivity.

Our main contributions of this paper are as follows.
1) We develop a new regression model for the dynamically-changing correlation matrices across all  time points and avoid regressing  in each correlation matrix separately.
2) The proposed method is applied to the  dynamic correlation matrices in resting-state fMRI, which are  further  partitioned  into disjoint brain states by the $k$-means clustering. 
3) We apply statistical tests on the dynamic connectivity states and transition matrices to  identify difference  between males and females in resting-state brain connectivity.

\section{Methods}

\subsection{Statistical model for dynamically-changing correlation matrices}
The dynamically-changing correlation matrices are originally computed by the sliding window method, which is defined as the Pearson correlation of consecutive windowed segments of fMRI time series \citep{keilholz2013dynamic,Hutchison2013}.
However, there are  unwanted high-frequency fluctuations and noise in the original dynamic correlation matrices. Thus, our goal is to produce the {\em smooth}  estimates of the dynamic correlation matrices.

Consider $p \times p$ dynamic functional connectivity such as correlation and covariance matrices obtained from fMRI time series in $p$ brain regions. The observed connectivity $C(t)$ at time $t$ is  modeled as
\begin{align*}
C(t)= \mu(t)+e(t),
\end{align*}
where $\mu(t)$ is the true underlying dynamic functional connectivity that has to be estimated, and $e(t)$ is noise. 

Let $Sym_p$ be the space
of all $p\times p$ 
symmetric matrices with inner product  $\langle A,B\rangle=\textmd{tr}(AB)$.
The space of $p\times p$ symmetric positive-definite (SPD) matrices \citep{arsigny2007geometric}, denoted by $Sym^+_p$, is a subspace of $Sym_p$. The exponential of a symmetric matrix is SPD, and the 
logarithm of  an SPD matrix is a symmetric matrix.
Moreover, the exponential map is one-to-one between $Sym_p$ and $Sym^+_p$ \citep{arsigny2007geometric,moakher2006symmetric}. Given  $X\in Sym_p$, its exponential map $\exp(X)$ is defined by matrix exponential  \citep{hall2015lie}
$$
\exp(X)=\sum_{n=0}^{\infty}\frac{X^n}{n!}.
$$
Let $I_{ij}$  be the  $p\times p$ matrix whose $(i, j)$-th and $(j,i)$-th entries  are $1/\sqrt{2}$ if
 $i\neq j$ and all  other entries are 0. Let $I_{ii}$ be the $p\times p$ diagonal matrix whose $(i, i)$-th entry is $1$ and all other entries are 0. For instance, for $p=3$, 
\begin{align}\label{eq:I3}
I_{31}=\frac{1}{\sqrt{2}}\begin{pmatrix} 
0 & 0 & 1 \\
0 & 0  & 0 \\
1 & 0  & 0 
\end{pmatrix} \textmd{ and } I_{22}=\begin{pmatrix} 
0 & 0 & 0\\
0 & 1  & 0 \\
0 & 0  & 0 
\end{pmatrix}.
\end{align}
Then,  we can show that $I_{ij}$ for $i\geq j$ form an orthonormal basis in $Sym_p$.

The matrix exponential is computed as follows. Suppose that the SPD matrix $X$ has factorization $X=UDU^{-1}$ where $D$ is the diagonal matrix with diagonal entries $d_i$. 
Then, the matrix exponential is computed as 
$$\exp(X)=U\exp(D)U^{-1}$$ 
where $\exp(D)$ is the diagonal matrix with diagonal entries given by $\exp(d_i)$ \citep{hall2015lie}.
For instance, the matrix  exponentials of  $I_{31}$ and $I_{22}$ in (\ref{eq:I3}) are 
\begin{align*} 
\exp(I_{31})
\approx\begin{pmatrix} 
    1.2606  &       0   & 0.7675\\
         0    & 1  &     0\\
    0.7675      &   0   &1.2606
\end{pmatrix} \textmd{ and } \exp(I_{22})=\begin{pmatrix} 
1 & 0 & 0\\
0 & \exp(1)  & 0 \\
0 & 0  & 1 
\end{pmatrix}.
\end{align*}

At fixed time point $t$, the $p\times p$ underlying dynamic connectivity $\mu(t)$  is estimated as  a linear combination of the $p\times p$ matrices $\exp (I_{ij})$,
\begin{align*}
\mu(t)=\sum_{1\leq j\leq i\leq p} b_{ij}(t)\exp (I_{ij}),
\end{align*}
where $b_{ij}(t)$ is the time-varying expansion coefficient. We further estimate coefficients $b_{ij}(t)$ using the Fourier cosine basis in time domain. The Fourier basis  has been widely used to reveal the spectral information of time series for further processing such as smoothing, regression and denoising. To simplify the problem, we restrict the time domain of $b_{ij}(t)$ to unit interval, i.e., $t\in[0,1]$,
by scaling the temporal resolution of fMRI  time series.
Then, $b_{ij}(t)$ can be represented by the linear combination of cosine basis, $1$ and  $\sqrt{2}\cos(\pi l t)$, and estimated as
\begin{align*}
b_{ij}(t)=b_{ij0}+\sqrt{2}\sum_{l=1}^{L}b_{ijl}\cos ({\pi lt}),
\end{align*}
where $b_{ijl}$ are  the cosine series coefficients estimated by the least squares method \citep{chung.2010.SII}.

\begin{figure}[!ht]
\centering
\includegraphics[width=1\linewidth]{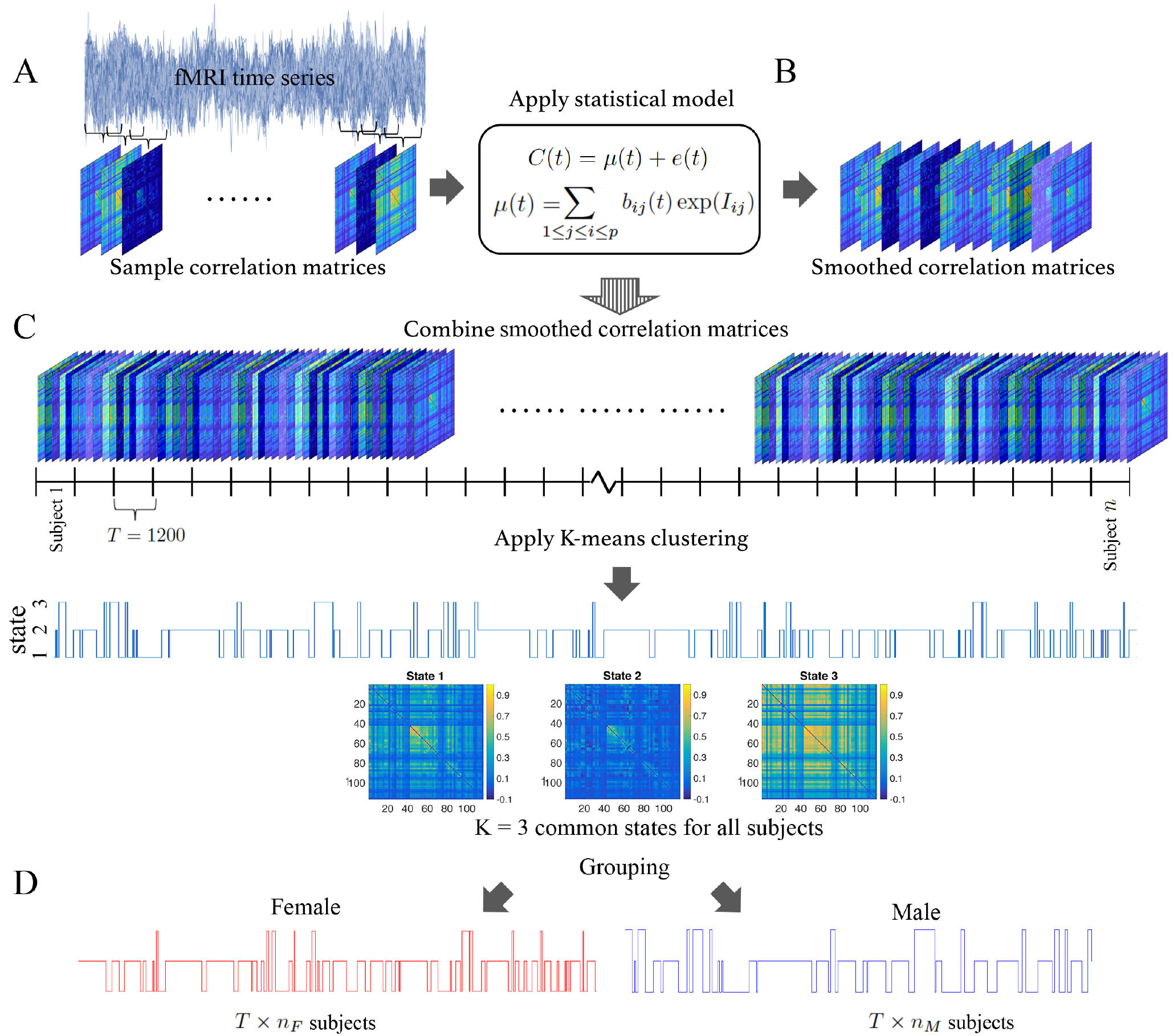}
\caption{
Schematic of estimation of 
dynamic connectivity  states.
A) For each subject,  $T$ sample 
correlation matrices are computed by the sliding-window method.
B) Original dynamic correlation matrices, composed of the $T$  sample correlation matrices, are smoothed by the proposed statistical model.
C) Smoothed dynamic correlation matrices of all 
$n$ subjects are combined and partitioned by the $k$-means clustering to find $K$ common states.
D) 
$T\times n$  estimated state labels are divided into male and female groups ($n_M$ males and $n_F$ females).}
\label{fig:Cluster_visual}
\end{figure}

\subsection{Clustering of the state space}
The estimated dynamic  functional connectivity is used in determining the state  space.
Assume there are $n$ subjects in the dataset. Let $C_i(t_j)$ denote the $p\times p$ dynamic  connectivity for the $i$-th subject at time point $t_j$. 
Let $\mathbf{d}_j^i$ denote the
vectorization of  the $\frac{p \times (p-1)}{2}$ elements in  the upper (or lower) triangular part of matrix $C_i(t_j)$. 
The collection of $\mathbf{d}_j^i$ over all time points and subjects is  fed into the $k$-means clustering in identifying the recurring brain connectivity states that are common across subjects \citep{Barber2018}.
The optimal number of cluster  $K$ is determined by the elbow method which  has been widely used in  previous studies \citep{Allen2014,Rashid2014,nomi2016dynamic,abrol2017replicability,lehmann2017assessing,Barber2018,ombao2018statistical}.
Figure \ref{fig:Cluster_visual} shows the schematic of the estimation of dynamic connectivity states.

\subsection{Validation}
We validated  the proposed method in a simulation.  The MATLAB code for the simulation below can be downloaded from \url{http://www.stat.wisc.edu/~mchung/statespaces}. The simulation was independently performed 100 times, and their average is reported here. We simulated the time series of 20 subjects. The data for each subject consists of time series from $p=20$ regions, and the length of the time series is $T=300$. Then we simulated three states for each subject as follows.

The state of each subject was uniformly chosen from 1, 2 and 3  with time duration of each state  randomly chosen from 5,  10, $\cdots$, 100 time points. Using  this random state space as the ground truth, we simulated time series for $p=20$ regions for each subject as follows. We started with generating five $5\times 1$ data vectors ${\bf x}_1, \cdots, {\bf x}_5 \sim {\cal N}(0, I)$,  identical and independently distributed multivariate normal with mean zero and identity matrix $I$ as the covariance. The data vector ${\bf x}_i$ is a noisy time series at 5 time points. Then the $5\times 1$ data vector ${\bf y}_i$ at  region $i$ was generated with dependency as follows. We  simulated ${\bf y}_1, \cdots, {\bf y}_4$ identical and independently distributed multivariate normal with
$${\bf y}_1, \cdots, {\bf y}_4 \sim  {\cal N}({\bf x}_1, \ 0.1^2 I).$$
Similarly we generate
\begin{align*}
{\bf y}_{5}, \cdots ,{\bf y}_{8}& \sim  {\cal N}( {\bf x}_{2} , \ 0.1^2 I)\\
&\ \ \vdots \nonumber\\
{\bf y}_{17}, \cdots ,{\bf y}_{20}& \sim {\cal N}({\bf x}_{5}, \ 0.1^2 I).
\end{align*}
The above simulation produces modular structures in the network \citep{chung.2017.CNI}.

Let ${\bf Y}_1$ be the $20\times 5$ data matrix 
$${\bf Y}_1 = [ {\bf y}_1, \cdots, {\bf y}_{20}]^{\top}.$$
Then correlation matrix $C^1=(c^1_{ij})$ corresponding to data matrix ${\bf Y}_1$ is given by
$c^1_{ij} = corr( {\bf y}_i, {\bf y}_j).$
We repeated the above procedure twice more  to  obtain three independent data matrices ${\bf Y}_1$, ${\bf Y}_2$ and ${\bf Y}_3$ and corresponding correlation matrices  $C^1$, $C^2$ and $C^3$ with spatial dependency between regions. The random correlation matrices with values close to each other may be difficult to  discriminate. Hence, we only used ${\bf Y}_k$ that will give the mean squared error (MSE) between  $C_1$, $C_2$ and $C_3$ larger than 0.5:
$$\frac{1}{20^2}\sum_{i,j} \left| c^{k_1}_{ij}- c^{k_2}_{ij} \right|^2>0.5 \ \ \mbox{ for all } k_1 \neq k_2.$$

We simulated the time series in $p=20$ regions for each subject as follows. If the subject is in state $k$ with time duration $5l$, the time series within this state are simulated by concatenating $20 \times 5$ data matrix $Y_k$ repeatedly $l$ times. For instance, if the subject is in state 2 with time duration 10,  state 1 with time duration 25,  followed by  state 3 with  time duration 5, the time series were simulated as
$${\bf Y} = [{\bf Y}_{2},{\bf Y}_{2},{\bf Y}_{1},{\bf Y}_{1},{\bf Y}_{1},{\bf Y}_{1},{\bf Y}_{1},{\bf Y}_{3}, \cdots],$$
where the size of whole data matrix ${\bf Y}$ is $20 \times 300$ representing time series at $p=20$ regions over 300 time points. We then added noise with standard deviation $\sigma$ in the range between 0.8 and 2.4 to make each block ${\bf Y}_k$ slightly different from each other. 

\begin{figure*}[t]
\centering
\includegraphics[width=1\linewidth]{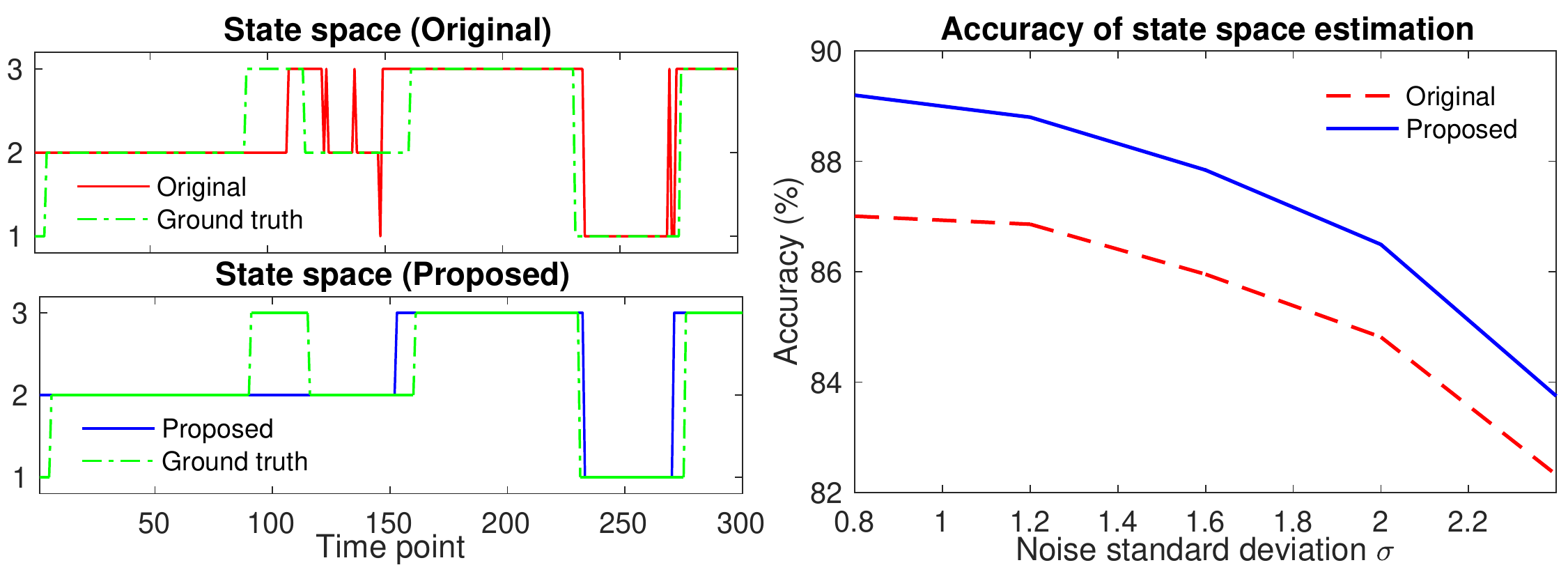}
\caption{Left: Estimates of state space of the original and the  proposed method compared to the ground truth.
We only showed the results of one subject in one simulation with noise standard deviation $\sigma=2$.
The proposed method has less rapid changes caused by noise.
Right: Accuracy of state space estimation for $\sigma=0.8$ to $2.4$ measured by the fraction of the estimated states equal to the ground truth.
The average of 100 independent simulations was plotted.}
\label{fig:SimAcc}
\end{figure*}

The above process is repeated 20 times for 20 subjects. We applied the sliding window method with window size 60 to obtain the original estimation of dynamic connectivity. We also applied the proposed method to smooth the dynamic connectivity. The proposed method has much less rapid state changes caused by noise (Figure~\ref{fig:SimAcc}-left), and thus has higher accuracy in state space estimation (Figure ~\ref{fig:SimAcc}-right).
\\

\section{Application to  resting-state fMRI}
\subsection{Dataset and preprocessing} 
The dataset is the resting-state fMRI of 412 subjects collected as part of the 
Washington University - University of Minnesota Human Connectome Project (HCP) \citep{vanessen.2012,van2013wu}. The resting-state fMRI were collected over 14 minutes and 33 seconds 
using a gradient-echo-planar imaging (EPI) sequence with multiband factor 8, time repetition (TR) 720 ms, time echo (TE) 33.1 ms, flip angle $52^\circ$, $104\times90$ (RO$\times$PE) matrix size, 72 slices, 2 mm isotropic voxels, and 1200 time points. 
During each scanning, participants were at rest with eyes open with relaxed fixation on a projected bright cross-hair on a dark background  \citep{van2013wu}.

The standard  minimal preprocessing pipelines \citep{glasser2013minimal} were applied on the fMRI scans including:
spatial  distortion removal \citep{jovicich2006reliability,andersson2003correct}, motion correction \citep{jenkinson2001global,jenkinson.2002},
bias field reduction \citep{glasser2011mapping}, 
registration to the structural MNI template,
and data masking using the brain mask obtained from FreeSurfer \citep{glasser2013minimal}.
The resulting volumetric data contains resting-state functional time series with $91\times109\times91=902629$ 2-mm isotropic voxels at 1200 imaging volumes.
\\

\noindent{\bf\em AAL parcellation.}
We employed the Automated Anatomical Labeling (AAL)  template to parcellate the brain volume into 116 non-overlapping anatomical regions \citep{tzourio.2002}. 
Spatial denoising was applied by averaging the fMRI data across voxels within each brain region, resulting in 116 average fMRI time series with 1200 time points for each subject. 
\\

\noindent{\bf\em Scrubbing.} Previous studies reported that  head movement produces spatially structured artifacts in functional connectivity \citep{power.2012,van2012influence,satterthwaite.2012,power2015recent,caballero2017methods}.
 Thus, scrubbing  was applied to remove fMRI volumes with significant head motion  \citep{power.2012}.
We calculated  the framewise displacement (FD)  from the three translational displacements ($x$, $y$, and $z$ axes) and three rotational displacements (pitch, yaw, and roll) at each time point \citep{power.2012} to measure  the head movement from one volume to the next.
The first volume of each subject is assumed to have zero FD.
To reduce the effect of head movement \citep{van2012influence}, the  volumes with FD larger than 0.5 mm and  their neighbors (one back and two forward) were scrubbed \citep{power.2012,power2013steps}.
We excluded 12 subjects having excessive head movement, and fMRI data of the remaining 400 subjects (168 males and 232 females) were used. More than one third of 1200 volumes being scrubbed in the excluded 12 subjects.
\\

\noindent{\bf\em Data imputation and bandpass filtering.} 
The imputation of the scrubbed data from the unscrubbed good data is often performed using the cubic spline interpolation in the previous studies \citep{Allen2014,Rashid2014,power2014methods,thompson2015frequency}. 
Further, to suppress the influence of low-frequency noise such as scanner drifts and high-frequency cardiac or respiratory oscillations \citep{cordes2001frequencies,van2008small}, temporal denoising  by bandpass filtering is also often  used \citep{muschelli2014reduction,thompson2015frequency,thompson2016bursty}.

In this paper, we performed data imputation and bandpass filtering together by the Fourier cosine basis \citep{lanczos1938trigonometric,hamming1998digital}
with cutoff frequencies of 0.01 and 0.1 Hz \citep{muschelli2014reduction,thompson2015frequency,thompson2016bursty}.

\subsection{Dynamic functional connectivity}
For each subject, we measured the  whole-brain dynamic functional connectivity by the $116\times116$ dynamic correlation matrix computed from the average fMRI signals in the 116 brain regions using the sliding window method.
\citet{shirer2012decoding,leonardi2015spurious} have reported that brain states may be correctly identified by a window size  in the range of  30--60 seconds.
In \citet{Allen2014}, window size 44 seconds was suggested to provide a good tradeoff between the ability to resolve dynamics and quality of
covariance matrix estimation.
Following  \citet{Allen2014}, we adopted window size 60 TRs, i.e., 43.2 seconds as TR=0.72 seconds in HCP dataset.
The sliding window is related to a smoother (low-pass filter) with  bandwidth $1/42.3=0.0236$ Hz.
To further smooth out the remaining high-frequency noise and fluctuations in the dynamic connectivity,
we performed the proposed statistical model with cosine series expansion  of degree 40 (corresponding to bandwidth $0.0236$ Hz). We then compared the proposed method with the original estimation of the dynamic connectivity.\\

\begin{figure}[t]
\centering
\includegraphics[width=0.7\linewidth]{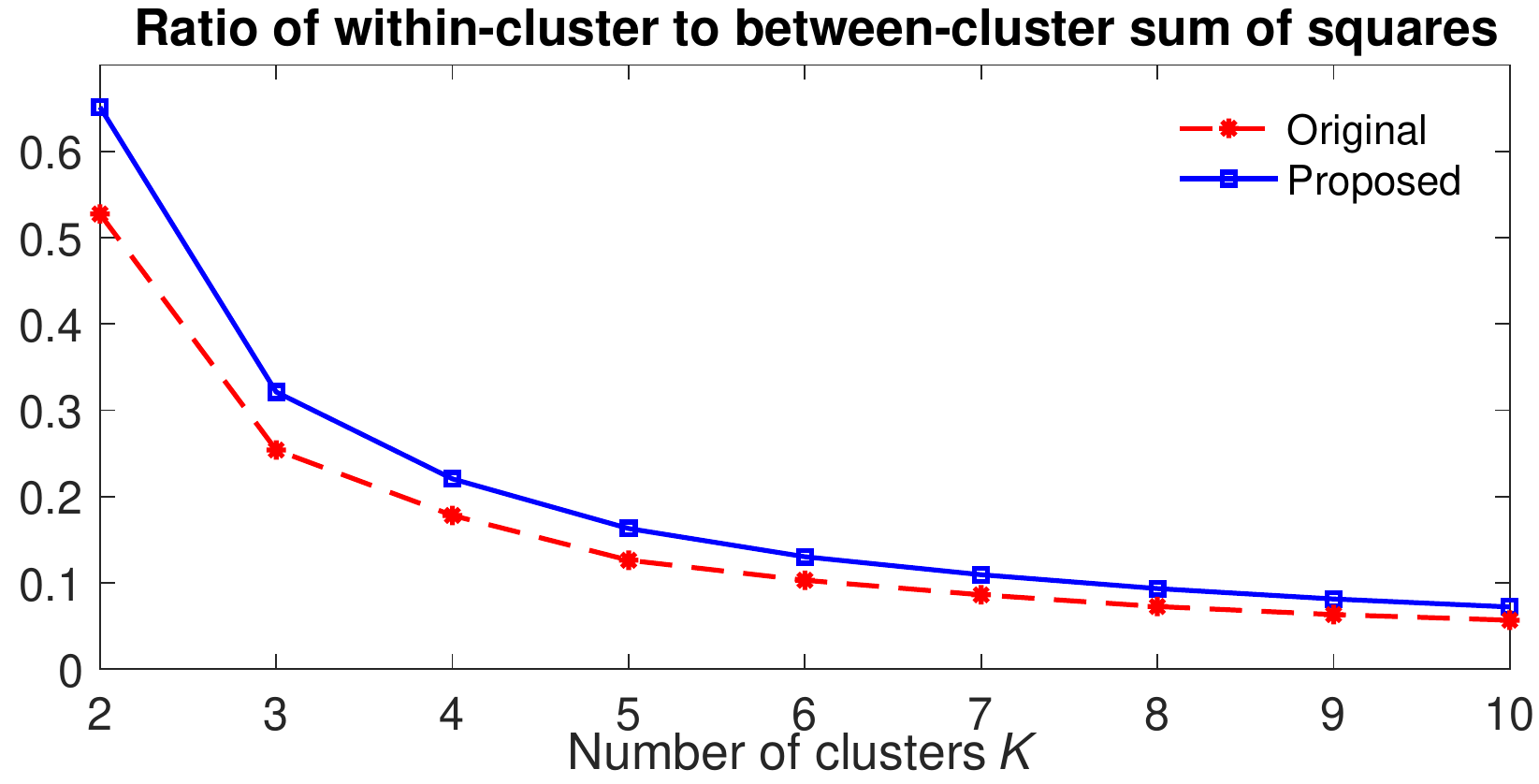}
\caption{The ratio of within-cluster to between-cluster sum of squared distances versus the number of clusters $K=2, \cdots,10$ for the original estimation and the proposed method.
By  the elbow method, $K=3$ was chosen since the slope changes the most drastically   at the elbow point $K=3$.}
\label{fig:elbow}
\end{figure}

\subsection{Dynamic connectivity states}
The baseline $k$-means clustering was used to identify the distinct states that repetitively occur throughout the time course and are common across subjects.
These discrete states serve as the basis of investigating  brain connectivity.
We determined the number of clusters $K$ by the elbow method which  has been widely used in literature \citep{Allen2014,Rashid2014,nomi2016dynamic,abrol2017replicability,lehmann2017assessing,Barber2018,ombao2018statistical}.
For each value of $K$, we computed the within-cluster  and between-cluster sums of squares, i.e., the sums of squared Euclidean distances between the cluster centroids and the data points within and outside the clusters.
Then, we plotted the ratio of within-cluster to between-cluster sum of squares for $K=2,...,10$ (Figure \ref{fig:elbow}).
By the elbow method, we chose $K=3$ which gives the largest slope change in the elbow plot.
Three states were also adopted by many previous studies \citep{choe2017comparing,Barber2018,ting2018estimating}.

\begin{figure*}[t]
\centering
\includegraphics[width=1\linewidth]{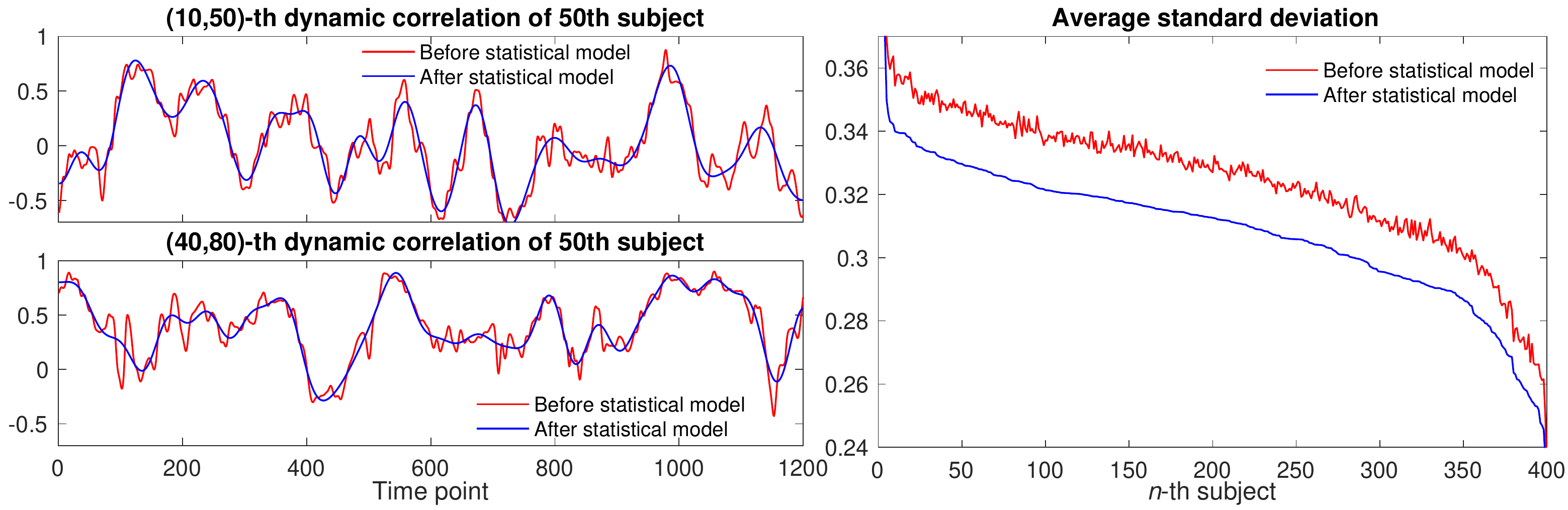}
\caption{Smoothness of the dynamic correlations by the proposed statistical model. 
Left:  The two entries of the dynamic correlation matrix of a subject.
Many high-frequency fluctuations were smoothed out by the proposed model.
Right: average standard deviations of the 400 subjects from averaging the standard deviations of the dynamic correlations  at the  6670 brain connections. They are sorted by the average standard deviations after the proposed model. The average standard deviations became smaller after smoothing by the proposed model.}
\label{fig:Smoothness}
\end{figure*}

\subsection{Results}
\noindent{\bf\em Variability in each subject.}
There are $116\times115/2=6670$  connections between the 116 brain regions.
Figure \ref{fig:Smoothness}-left shows the dynamic correlations of one subject at two connections.
Many high-frequency fluctuations and noise in the dynamic correlations were smoothed out by the proposed statistical model.
For each connection, we computed the standard deviation of the dynamic correlations over time. 
Then, we averaged the standard deviations across all  6670 connections.
The  average standard deviations of the 400 subjects are displayed in Figure \ref{fig:Smoothness}-right.
\\

\begin{figure*}[!ht]
\centering
\includegraphics[width=\linewidth]{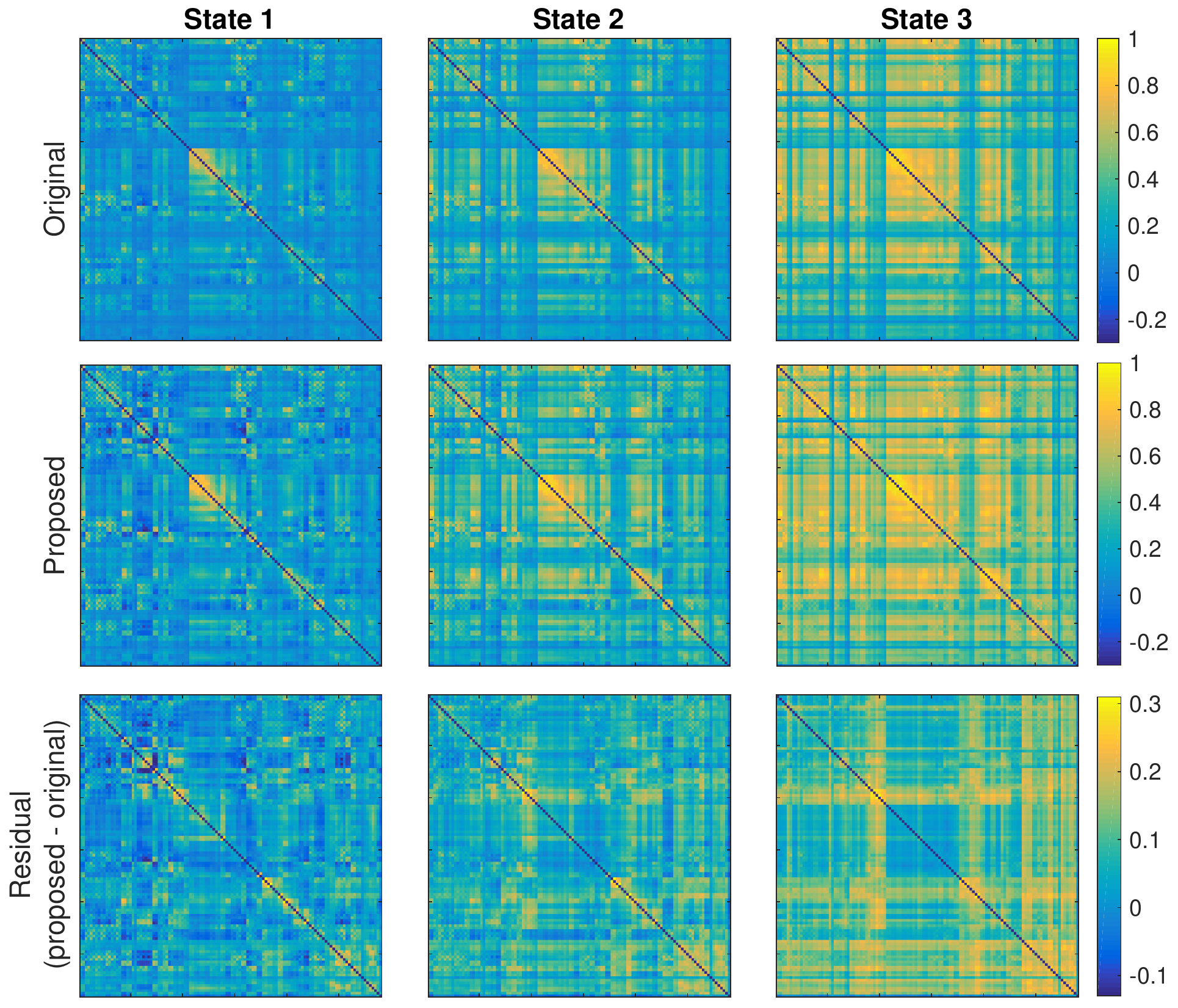}
\caption{Top and middle: average correlation matrices (cluster centroids) of the three states for the original estimation and the proposed method.
The minimum and maximum average correlations of the three states are $(-0.22,0.78)$, $(-0.06, 0.88)$ and $(-0.04, 0.93)$ for the original estimation,
and are $(-0.34,0.88)$, $(-0.15, 0.93)$ and $(-0.09, 0.96)$ for the proposed method.
Bottom: residual of the  average correlations (proposed - original) with  minimum and maximum given by $(-0.14,0.31)$,  
$(-0.09,0.31)$ and $(-0.05,0.31)$ in the three states.}
\label{fig:AvgCorr}
\end{figure*}

\begin{figure*}[!ht]
\centering
\includegraphics[width=\linewidth]{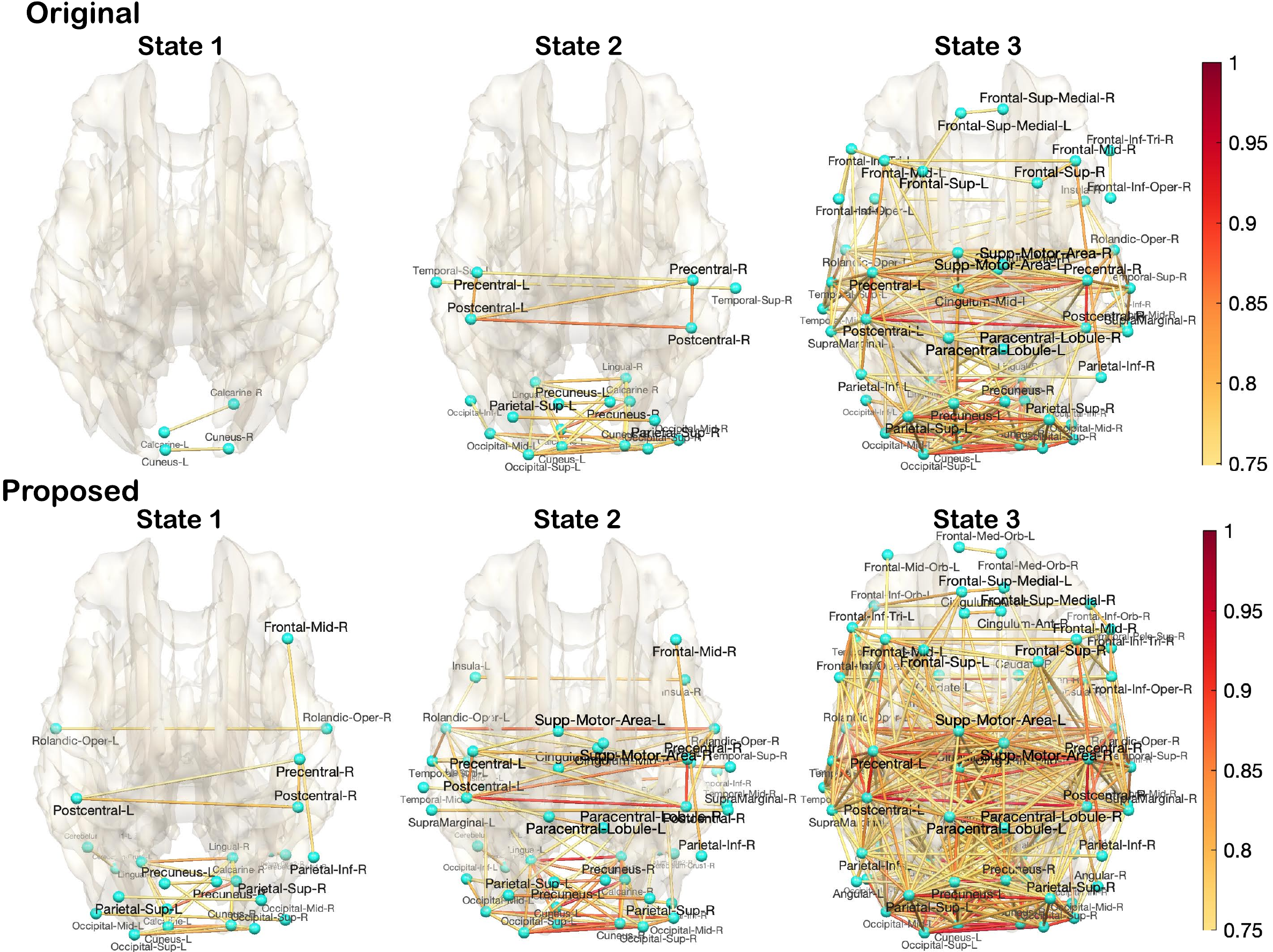}
\caption{Average correlation matrices (cluster centroids) of the three states for the original estimation (top) and proposed method (bottom).
Only connections with correlation larger than 0.75 are displayed.
For both methods, the four most connected regions in states 1 and 2 are within the occipital lobe, such as the calcarine fissure and surrounding cortex, cuneus and lingual gyrus. In state 3, besides the occipital lobe, the most connected regions also include the precentral gyrus, superior temporal gyrus, and median cingulate and paracingulate gyri among other regions.}
\label{fig:AvgCorrBrain}
\end{figure*}

\noindent{\bf\em Average correlation matrix within each state.} 
We applied the $k$-means clustering to the dynamic correlation matrices obtained from the original estimation and the proposed method. Figure \ref{fig:AvgCorr} shows the state-specific average correlation matrices, i.e, the cluster centroids. The proposed method shows a wider range of average correlations. The minimum and maximum average correlations of the three states are  $(-0.22,0.78)$, $(-0.06, 0.88)$ and $(-0.04, 0.93)$ for the original estimation,
and are $(-0.34,0.88)$, $(-0.15, 0.93)$ and $(-0.09, 0.96)$ for the proposed method.
The residual of the average correlations (proposed - original) ranges from $-0.14$ to $0.31$,
from $-0.09$ to $0.31$, and from $-0.05$ to $0.31$ in the three states.

Figure \ref{fig:AvgCorrBrain} is an alternative visualization of the average correlation matrices, showing strong connections with  average correlation larger than 0.75.
For the original estimation and proposed method, the
four most connected regions in states 1 and 2 are within the occipital lobe, such as the calcarine fissure and surrounding cortex, cuneus and lingual gyrus. In state 3, besides the occipital lobe, the most connected regions also include the precentral gyrus, superior temporal gyrus, and median cingulate and paracingulate gyri among other regions.
\\

\begin{figure}[t]
\centering
\includegraphics[width=1\linewidth]{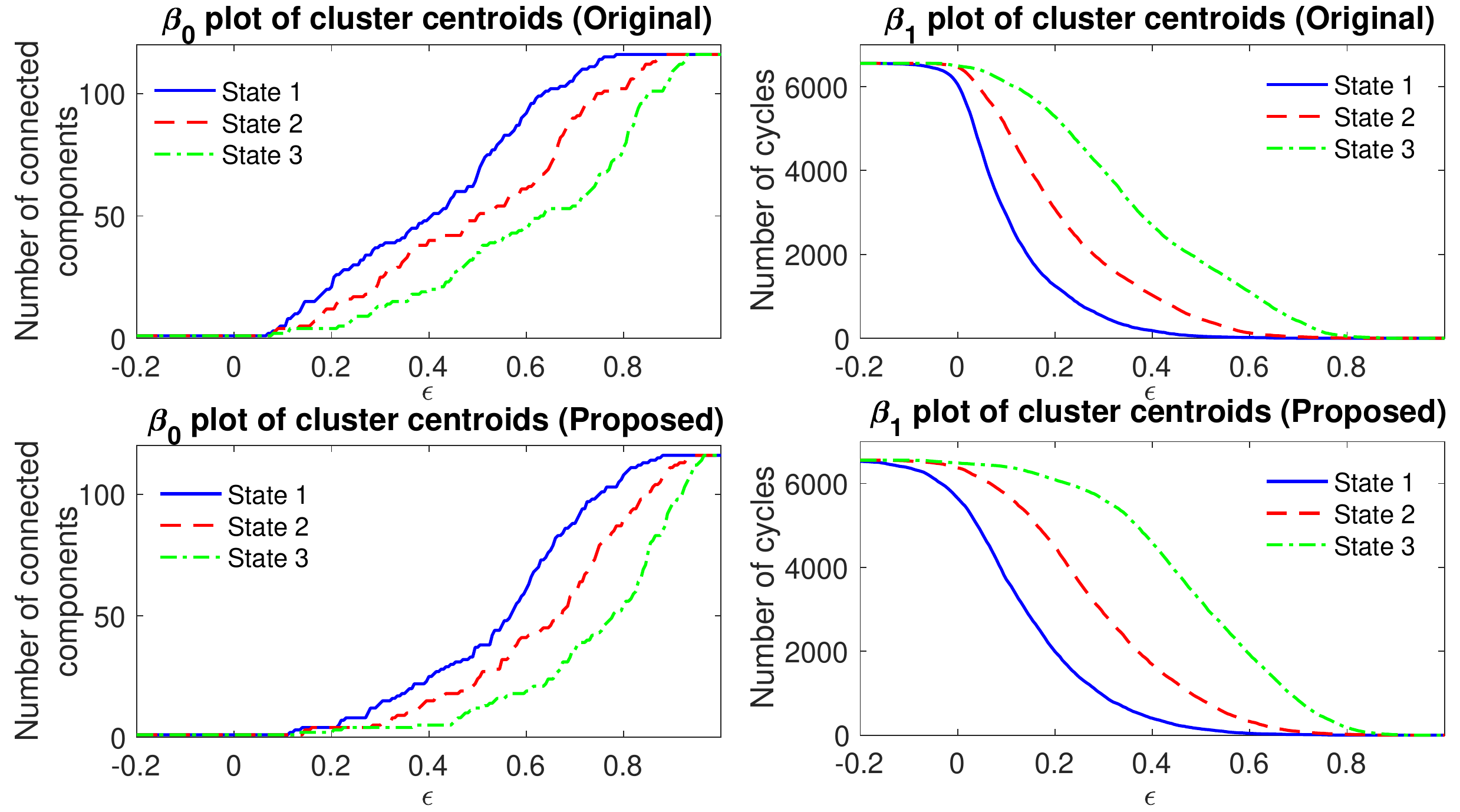}
\caption{
Betti numbers $\beta_0$ and $\beta_1$ of the brain networks thresholded by correlation value $\epsilon$. The edge weights of the networks are given by the average correlation matrices (cluster centroids) of
 the original estimation (top) and proposed method (bottom).
The $\beta_0$ differences between any two states are all  larger than 34 with $p$-values smaller than $9.4\times 10^{-5}$.
The $\beta_1$ differences between any two states are all  larger than 2152 with $p$-values smaller than $10^{-16}$.
In both the original estimation and proposed method, the connectivity patterns of the three states are all topographically different.}
\label{fig:betti}
\end{figure}

\noindent{\bf\em Topological  differences between states.}
We determined if the three estimated states are topologically different using the Exact Topological Inference  \citep{chung.2017.IPMI, chung.2017.CNI,chung2019statistical}. We computed the 0th Betti number $\beta_0$ and 1st Betti number $\beta_1$ of the brain network  in each state. We built brain networks with edge weights being the average correlations of each state. By thresholding   the edge weights at higher correlation value,  more edges in the networks  were removed, and hence the number of connected components ($\beta_0$)  increased while the number of cycles ($\beta_1$)  decreased. We used  threshold values  ranging from $-0.2$  to 1 at  an increment of 0.005.
The $\beta_0$- and $\beta_1$-plots of the average correlation matrices (cluster centroids) of the three states  are displayed in Figure \ref{fig:betti}. For the original estimation and proposed method, the  $\beta_0$ differences between any two states are all  larger than 34 with $p$-values smaller than $9.4\times 10^{-5}$.
The $\beta_1$ differences between any two states are all  larger than 2152 with $p$-values smaller than $10^{-16}$.
The Bonferroni correction \citep{bonferroni1936teoria,shaffer1995multiple} rejects the null hypothesis that the three states are  topologically equivalent at a significance level of $\alpha=0.01$.
\\

\begin{figure*}[t]
\centering
\includegraphics[width=\linewidth]{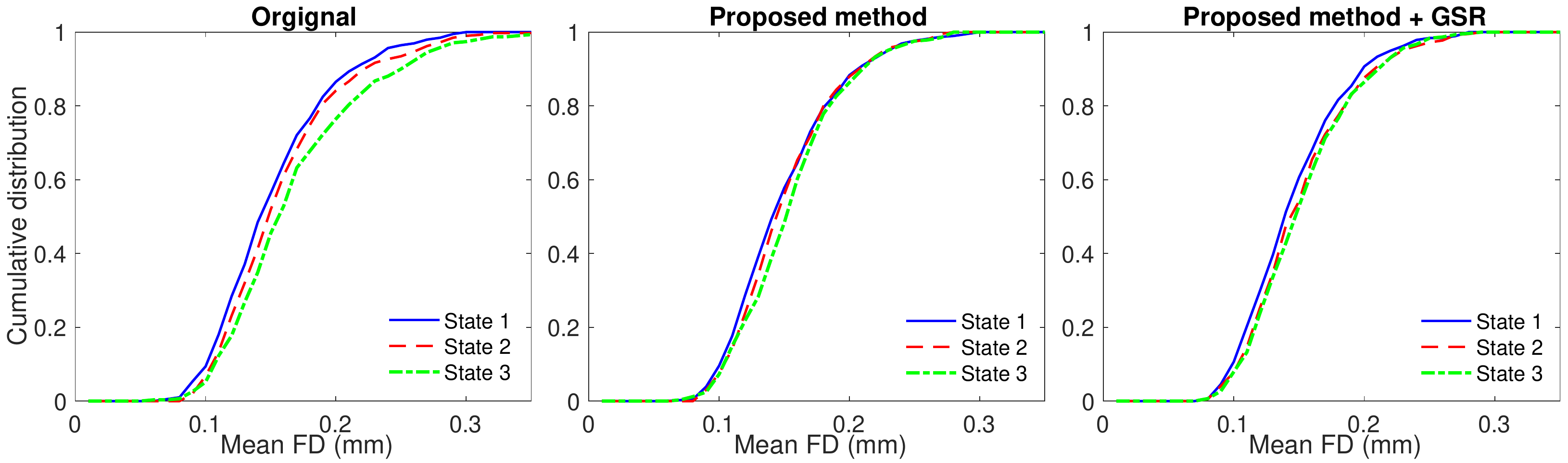}
\caption{Cumulative distribution function of the mean framewise displacement (FD) of  all subjects at each state, where the state spaces were  obtained from the original estimation (left),  proposed method (middle), and proposed method plus GSR (right).
We compared the distributions of the three states by the Kolmogorov-Smirnov test with the Bonferroni correction. At  a significance level of  $\alpha=0.01$,  there are state differences in the original estimation ($p=4.7\times10^{-4}$) but no state differences in the proposed method either with or without GSR ($p=0.0326$, 0.1570).}
\label{fig:FD}
\end{figure*}

\noindent{\bf \em Framewise displacement within each state.}
As mentioned previously, the framewise displacement (FD) was used to measure the head movement.
For each subject, we computed the mean FD within each state. 
Figure \ref{fig:FD} shows the distributions (cumulative distribution functions) of the mean FD of all subjects in the three states, where the state spaces were  obtained from the original estimation, the proposed method, and the proposed method plus global signal regression (GSR) \citep{murphy2009impact}.
We performed the Kolmogorov-Smirnov test \citep{massey1951kolmogorov} with the Bonferroni correction to compare the  distributions of the three states.  For the three methods,  the $p$-values are  $4.7\times10^{-4}$,  0.0326 and 0.1570 respectively. Thus, at a significance level of  $\alpha=0.01$,
there  are state differences in head movement in the original estimation but no state differences in head movement in the proposed method with or without GSR. 
Thus, GSR was not used in this study.
\\

\begin{figure*}[!ht]
\subfloat[][Original estimation]{\centering
\includegraphics[width=\linewidth]{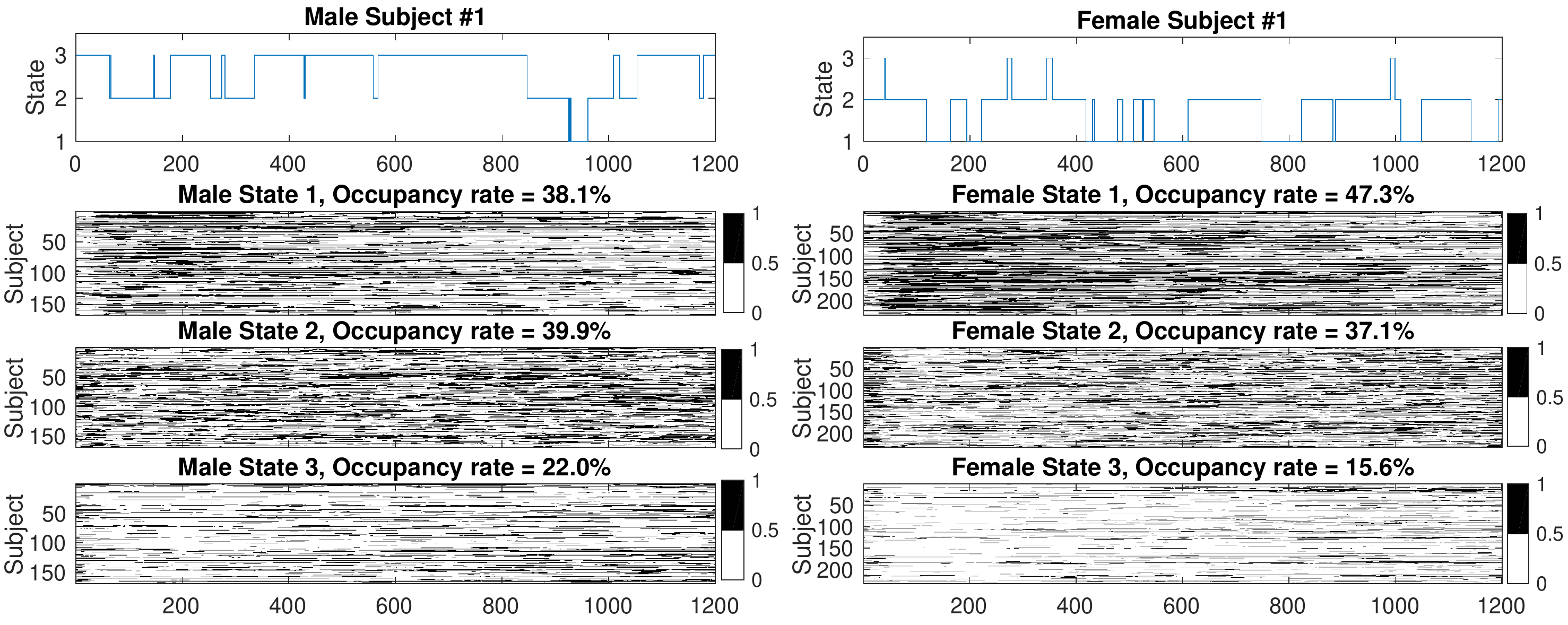}}\\
\subfloat[][Proposed method]{\centering
\includegraphics[width=\linewidth]{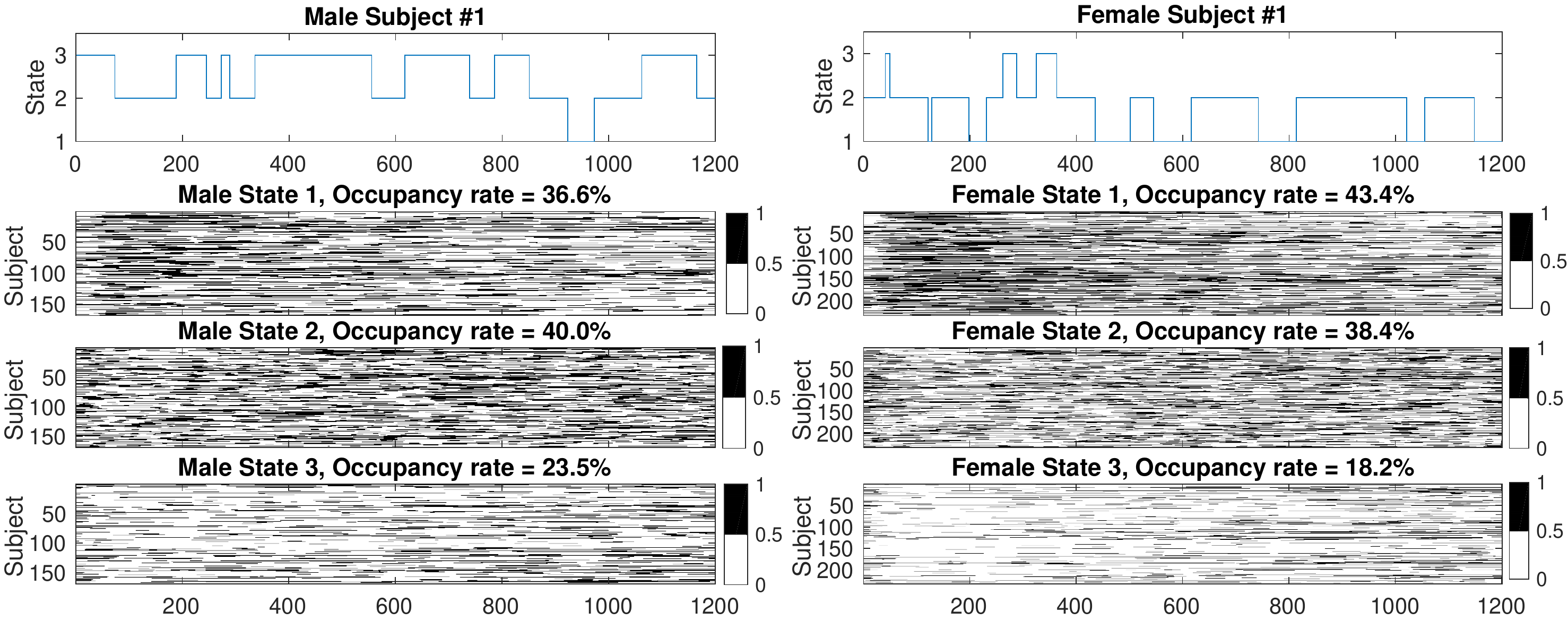}}
\caption{In the (a) original estimation and (b) proposed method, the first row shows the dynamic connectivity 
states of the 1st male (left) and 1st female (right) subjects.
The proposed method  has less rapid changes  in the  dynamic connectivity states. 
The 2nd to 4th rows are the plots of he state occupancy of all the 168 males (left) and 232 females (right).
Correlation matrices belonging to the state are marked by black.}
\label{fig:BinOccupant}
\end{figure*}

 \begin{figure*}[!ht]
\centering
\includegraphics[width=1\linewidth]{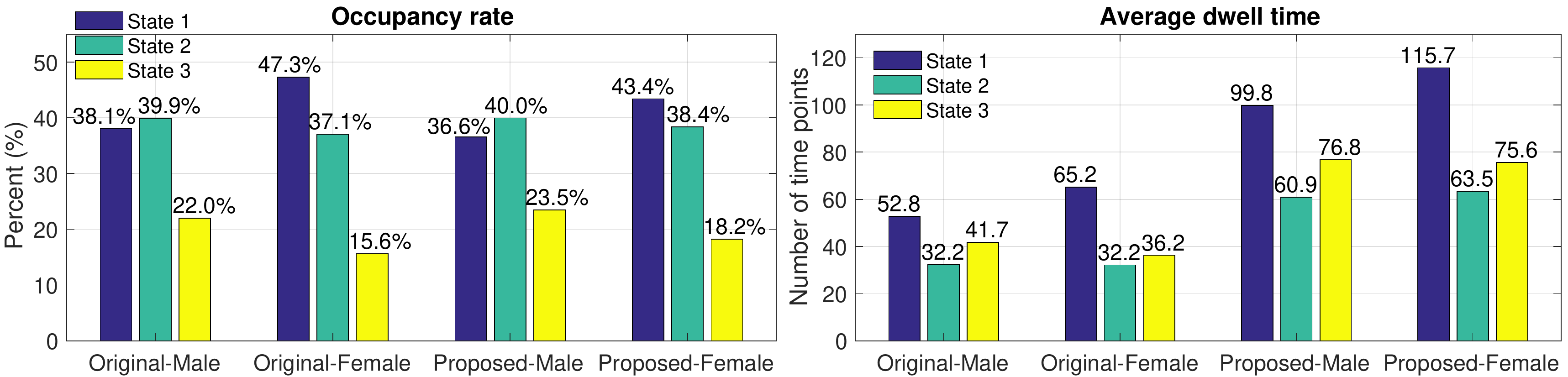}
\caption{Left: Occupancy rates of the three states, i.e., the percentage of the entire scan time that a male/female spends in each state.
On average, males spent more time in state 2 while females spent more time in state 1.
Right: Average dwell time of the three states,  i.e., the average period of time a male/female remains in a given state before switching to another state. On average, a subject dwelt in state 1 for a longer period, and females had longer dwell time in state 1 than males. The proposed method has longer average dwell time than the original estimation due to less rapid state changes.}
\label{fig:Occupancy_Dwell}
\end{figure*}

\noindent{\bf \em Occupancy rate and dwell time.}
To analyze the difference between males and females in dynamic connectivity states, we further divided the 
clustering results into male and female groups (168 males and 232 females).
Figure \ref{fig:BinOccupant} shows the 
dynamic connectivity states of the 1st male and 1st female subjects as an example. The proposed method has less number of rapid changes  in the  dynamic connectivity states.
Figure \ref{fig:BinOccupant} also shows the  occupancy of the three states for all male and female subjects.
Let  $s_{ij}\in\{1,2,3\}$ denote the state of subject $i$ at time point $t_j$ estimated  by the $k$-means clustering.
The occupancy rate \citep{yaesoubi2015dynamic,ombao2018statistical} of state $k$ is computed as
$$ \frac{1}{nT} \sum_{i=1}^n\sum_{j=1}^{T} \left(s_{ij} = k\right),$$
where $T=1200$ time points and $n=168$ and 232 subjects for male and female groups respectively.
On average, males spent more time in state 2, while females spent more time in state 1 (Figure \ref{fig:Occupancy_Dwell}-left).

We also computed the dwell time \citep{Damaraju2014,lottman2017risperidone,Barber2018}, i.e., the period of time a male/female remains in a given state before switching to another state.
The  average dwell time  for males and females in each state  is displayed in Figure \ref{fig:Occupancy_Dwell}-right.
On average, a subject dwelt in state 1 for a longer period, and females had longer dwell time in state 1 than males. The proposed method shows longer average dwell time than the original estimation due to less rapid state changes.
\\

\begin{figure*}[!ht]
\centering
\includegraphics[width=\linewidth]{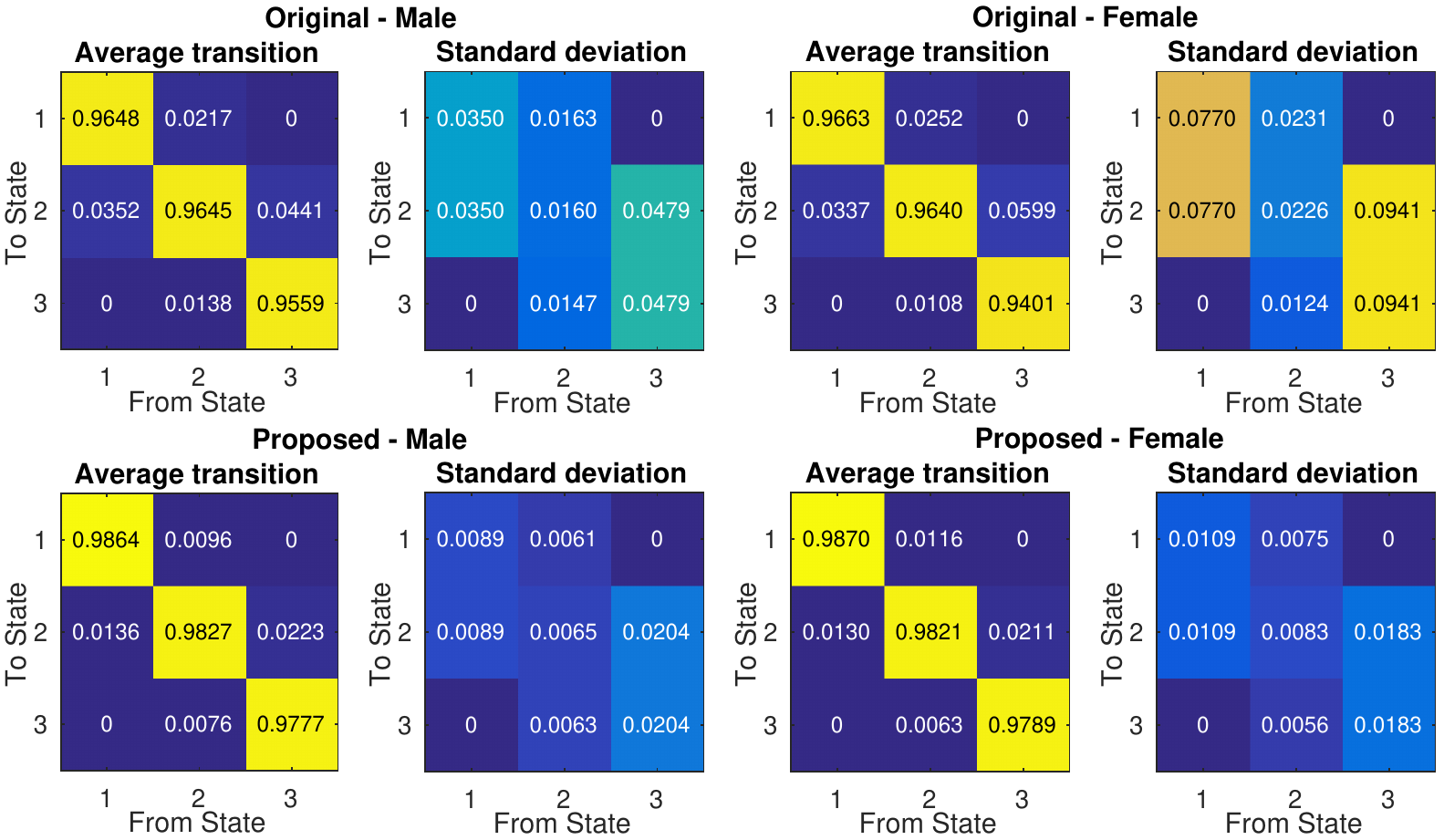}
\caption{Averages and standard deviations of transition probabilities of males (left) and females (right) computed from the dynamic connectivity states of the original estimation (top) and the proposed method (bottom).
The proposed method reduced the transition probabilities between different states and increased the probabilities of remaining in the same state.}
\label{fig:Transition}
\end{figure*}

\noindent{\bf \em Transition probability.} 
We  used state transitions to reveal the  interactions between different brain states \citep{baker2014fast}.
They can be modeled as a Markov chain \citep{gilks1995markov}.
For  subject $i$,  the transition probability of moving from state $k_1$ to state $k_2$ is computed by 
$$ P\left(s_{ij}=k_2|\ s_{i,j-1}=k_1\right),$$
where $s_{ij}$ is the state of subject $i$ at time point $t_j$ estimated by the $k$-means clustering.
Figure \ref{fig:Transition} shows the averages and standard deviations of the transition probabilities 
of males and females.
Each subject remained in the same state for a long period of time before transitioning to other  state. The proposed method reduced the transition probabilities between different states and increased the probabilities of remaining in the same state because some transitions caused by noise were removed.
The very low average transition probabilities between state 1 and state 3 show the inability of transitioning directly between  these two states.	
\\

\begin{figure*}[t]
\centering
\includegraphics[width=\linewidth]{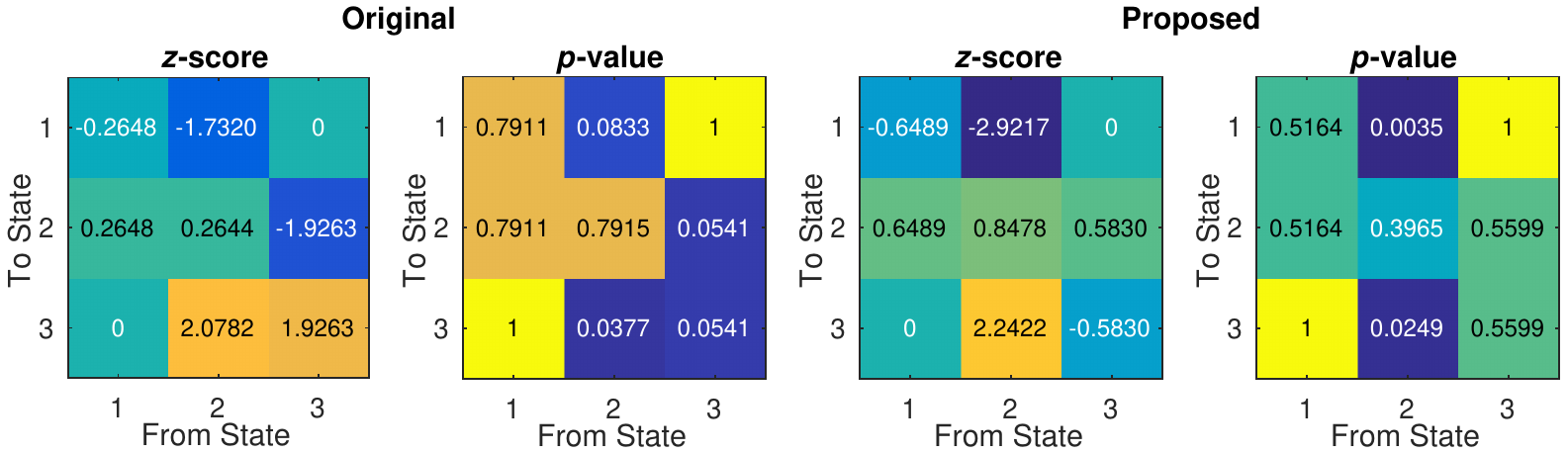}
\caption{Z-test for differences in transition probabilities between 
males and females using  the difference in the mean proportions at significance level 0.05.
In the original estimation (left), there is significant difference in the transition probability of  ($2 \rightarrow 3$).
In the proposed method (right), there are significant differences in the transition probabilities of ($2 \rightarrow 1$) and ($2 \rightarrow 3$).}
\label{fig:Zscore}
\end{figure*}

\noindent{\bf \em Statistical analysis.}
To compare  males and females in state transition probability, comparison of 
mean sample proportions was utilized.
Consider the null hypothesis that the  averages of estimated transition probabilities from state $k_1$ to $k_2$
for males and females are the same. 
The z-score was computed from
\begin{equation*}
Z = \frac{\bar{P}_{M} - \bar{P}_{F}}{\sqrt{ \frac{\sigma^2_{M}}{n_M} + \frac{\sigma^2_{F}}{n_F} }},
\end{equation*}
where $\overline{P}_{M}$ and $\overline{P}_{F}$ are the means of the transition probabilities for  $n_M$ males and  $n_F$ females respectively, 
and $\sigma^2_{M}$ and $\sigma^2_{F}$ are the variances of the transition probabilities.
The z-scores and the corresponding $p$-values are shown in Figure \ref{fig:Zscore}. 
In the original estimation, there is significant difference at $p$-values $<0.05$ in the transition probability from state 2 to state 3  ($2 \rightarrow 3$).
In the proposed method, there are significant differences at $p$-values $<0.05$ in the transition probabilities of ($2 \rightarrow 1$) and ($2 \rightarrow 3$).

We further tested the statistical significance of state occupancy rates between females and males, by setting the null hypothesis 
that males and females have the same mean of occupancy rate.
The $p$-values of the  z-test are shown in Table \ref{table:1}. 
For both the original estimation and proposed method, there are differences in the occupancy rates of states 1 and 3 at significance level 0.05.

\begin{table*}[t]
\caption{$p$-values of the z-test for gender differences in state occupancy rates.} 
\label{table:1}
\centering
\begin{tabular}{ | l | l | l |  }
\hline
State & Original & Proposed   \\ 
\hline
1 & $0.0019$ &  $ 0.0114$ \\ 
\hline
2& $0.1445$ &  $0.3685$  \\ 
\hline
3 & $0.0050$ & $0.0215$  \\
\hline
\end{tabular}
\end{table*}

\section{Discussions}
\noindent{\bf\em Dynamic connectivity.}
In this study, we found that the average correlation matrices (cluster centroids) of the three states followed similar connectivity patterns of the previous study \citep{haimovici2017wakefulness},  which also used the AAL parcellation but  $k$-means clustering with four states.
In \citet{haimovici2017wakefulness}, two of the four states show high average correlation in many brain connections.
In our result, the three states are not disjoint  but share similar connectivity pattern. This may be due to the small number of clusters, which can also be observed in \citet{cai2018estimation}.
Consider the resting state networks \citep{ting2018multi,al2019tensor}.
The average correlation matrices of the three states  show relative higher correlations in the occipital lobe, such as calcarine fissure and surrounding cortex, cuneus and lingual gyrus.
Compared to other resting state networks, the visual network has the strongest connectivity across different states, followed by the somatomotor network including brain regions such as the postcentral gyrus and precentral gyrus.
\\

\noindent{\bf\em  Transition probability of brain state.}
The resting-state networks tend to remain in the same state for a long period before switching to another state \citep{Allen2014,Shakil2016,calhoun2016time,abrol2017replicability,nielsen2018predictive}.
In this study, we  showed that the state space of the proposed method had a longer stability (less rapid changes and longer dwell time)  and a higher probability of remaining in the same state compared to the original estimation.
\\

\noindent{\bf\em Estimation of dynamic  functional  connectivity.}
The proposed model aims for smoothing out unwanted high-frequency fluctuations in the original estimation of dynamic connectivity which may introduce rapid changes in brain state estimation in resting state. 
There are a variety of  dynamic connectivity estimation methods besides the sliding window method, such as the tapered sliding window  \citep{Allen2014,lindquist2014evaluating,abrol2017replicability}, flexible least squares \citep{liao2014dynamicbc}, multiplication
of temporal derivatives \citep{shine2015estimation}, and jackknife correlation \citep{thompson2018simulations,thompson2018common}.
It is still unclear which method is the optimal since the true  dynamic connectivity is unknown.
For instance, \citet{thompson2018simulations} showed in simulations that the jackknife correlation outperforms the sliding and tapered sliding window methods when the state changes quickly, but the taped sliding window method followed sliding window method performs the best when the state changes slowly.
In this paper, we used the sliding window method as it may be the simplest and most widely used method, but  
the proposed model can be applied to other estimation methods.
The performance of the proposed model would vary depending on the original estimation of the dynamic connectivity.
\\

\noindent{\bf\em Smoothing of dynamic  functional connectivity.}
The proposed model contains a temporal smoothing, which fits the time-varying coefficients to a cosine series representation and estimates the cosine series coefficients by the least squares method \citep{chung.2010.SII}.
Other least squares smoothing methods can be applied instead, but the smoothing result might be different. For example, the least squares smoothing in  \citet{selesnick2012polynomial,baek2015baseline} minimizes the second-order difference to force the signal to be smooth. The Savitzky--Golay Filter \citep{savitzky1964smoothing,madden1978comments} fits the subsets of adjacent data points to a polynomial by linear least squares. This is left as a future study.

\section{Conclusion}

In the proposed regression method, the dynamic correlation matrix is modeled as a linear combination of symmetric positive-definite (SPD) matrices combined with cosine series representation, which provides superior performance over existing sample 
correlation matrices. We represented the correlation matrix, at each time point, as a linear combination of exponential map of the orthonormal basis in the space of symmetric matrices. Doing so, we smoothed out the unwanted noise in dynamic functional connectivity and achieved higher accuracy in identifying and discriminating brain connectivity states. 

\section*{Acknowledgements}
This study was supported by 
NIH Brain Initiative grant EB022856, NIH grant R01-MH11569 and KAUST. We would like to thank  Andrey Gritsenko, Gregory Kirk and Rasmus M. Birn of University of Wisconsin-Madison and Martin Lindquist of Johns Hopkins University for valuable discussions and logistic supports.

\section*{References}
\bibliographystyle{elsarticle-harv} 
\bibliography{reference.NN.2019.10.27}

\end{document}